\documentclass[12pt]{elsarticle}
\usepackage{amsmath}
\usepackage{amssymb}
\usepackage{color, a4wide}
\usepackage{epstopdf}

\biboptions{sort&compress}

\journal{Journal of Theoretical Biology}
\date{}

\begin{document}

\begin{frontmatter}

\title{A Condition for Cooperation in a Game on Complex Networks}
\author{Tomohiko Konno}
\address{Institute for Advanced Study, Waseda University}
\address{Totsuka-cho 1-104 Shinjuku-ku Tokyo 169-8050, Japan}

\ead{tomo.konno@gmail.com}

\begin{abstract}
We study a condition of favoring cooperation in a Prisoner's Dilemma game on complex networks. There are two kinds of players: cooperators and defectors. Cooperators pay a benefit $b$ to their neighbors at a cost $c$, whereas defectors only receive the benefit $b$. The game is a death-birth process with weak selection.  Although it has been widely thought that $b/c>\langle k \rangle$ is a condition of favoring cooperation \cite{Ohtsuki.Nature.2006}, we find that $b/c>\langle k_\textrm{nn} \rangle$ is the condition. We also show that among three representative networks, namely, regular, random, and scale-free, a regular network favors cooperation the most, whereas a scale-free network favors cooperation the least. In an ideal scale-free network where network size is infinite, cooperation is never realized. Whether or not the scale-free network and network heterogeneity favor cooperation depends on the details of a game, although it is occasionally believed that scale-free networks favor cooperation irrespective of game structures. If the number of players are small, then the cooperation is favored in scale-free networks.
\end{abstract}
\begin{keyword}
Games\sep Cooperation\sep Prisoner's Dilemma \sep Complex Networks
\end{keyword}

\end{frontmatter}

\section{Introduction}
 Although a player incurs a lot of cost for cooperative behavior and being selfish is usually more beneficial than being cooperative as one player, cooperative behavior is ubiquitously observed in various forms of life system including even single cell. Cooperation is even the basis of life system, eco-system, and animal society including human being. Thereby, the research on how cooperation emerges and being enhanced have attracted much attention \cite{nowak2006fre} in game theory. To answer the question Prisoner's Dilemma game is often used, in which being defector is always better off than being cooperator and, however, both of the players being cooperators are always better off than both of them being defectors. This game structure reflects the situation of interest in reality.

 The seminal paper~\cite{nowak.1992.nature} introduced a spatial structure in the game theory and showed that a lattice structure enhanced cooperation in Prisoner's Dilemma game. Since then, it has been well recognized that the spatial structure is one factor affecting the emergence of cooperation and thus a lot of effort has been made in this direction \cite{hofbauer.1998,nowak.2004.nature,killing.1996,nakamura.1997.theo,baalen.1998,taylor.2007.nature,mitterldorf.2000.theo,
ifti.2004.theo,lieverman.2005.nature,szabo.2004.pre,wilson.1992.evol,Masuda07082007,morita2008extended}. In contrast, cooperation is often inhibited by spatial structure in snow drift games \cite{hauert.2004.nature}. The spatial structure is regarded as a network. Many networks in reality are not often regular lattice; rather, they are often small-world; scale-free; or heterogenous heterogeneous networks. It has also been recognized that underlying network structures crucially determine an outcome of the models such as percolation, synchronization, epidemic spread, the Ising , voter, and a lot of other models~\cite{dorogovtsev.2008.rev,newman2003sfc}. Thereby, many researchers have expressed interest in games on such complex networks recently (see review~\cite{szabo.rev.2007}) and explored especially how network structures such as the small-world characteristics, the scale-freeness, the network heterogeneity, and so on, affect the emergence of cooperation. For instance, Refs.~\cite{santos2005scale,santos2008social,santos2006evolutionary} showed that the scale-free network enhanced cooperation.
Contrary to them, in the present paper we will conclude for the Prisoner's Dilemma model defined in section~\ref{sec: model} that cooperation is inhibited by scale-free or heterogeneous network and enhanced by regular network, agreeing with Refs.~\cite{nowak.1992.nature,fu2009evolutionary}.

The review~\cite{nowak2006fre} entitled ``Five rules for the evolution of cooperation'' listed five mechanisms for the evolution of cooperation: kin selection, direct reciprocity, indirect reciprocity, network reciprocity, and group selection. We will show that the condition of favoring cooperation exactly corresponds to the network reciprocity of the five mechanisms. More specifically, the condition is $b/c>\langle k_\textrm{nn}\rangle$ for general uncorrelated networks (in which the degree and the nearest neighbor degree do not correlate), where $b$ and $c$ are the benefit and the cost of the Prisoner's Dilemma game, and $\langle k_\textrm{nn}\rangle$ is the mean degree of the nearest neighbors. Although many preceding researches are numerical simulations, we will analytically derive the condition by using pair approximation and mean-field approximation in heterogenous networks.

Previously, however, it was widely believed for the same model that the general condition for favoring cooperation is $b/c>\langle k \rangle$ \cite{Ohtsuki.Nature.2006} (and reviews~\cite{nowak2006fre,szabo.rev.2007}), where $\langle k\rangle$ is the mean degree. The point here is the difference between $\langle k_\textrm{nn}\rangle$ and $\langle k \rangle$. The difference is essential in network theory, producing interesting results in complex networks; see Refs.~\cite{PhysRevLett.85.4626,PhysRevLett.86.3200,nature.406.378}. It is not too much to say that owing to the difference network theory can exist.
%
If a network has no degree-degree correlation, then $\langle k_\textrm{nn} \rangle=\langle k^2 \rangle/\langle k \rangle$.
The probability that an end of a link is attached to a vertex with degree $k$, $P_\textrm{nn}(k)$, is given by
\begin{align}
P_\textrm{nn}(k)&=\frac{\text{\# Ends attached to vertices with degree $k$}}{\text{\# All the ends of links in the network}}
\notag\\
&=\frac{N k P(k)}{N\sum_k k P(k)},
\end{align}
where $P(k)$ is the degree distribution.
Therefore, we have
\begin{align}
\langle k_\textrm{nn}\rangle &\equiv \sum_{k} k P_\textrm{nn}(k)\notag \\
&=\frac{\sum_k k^2 P(k)}{\sum_{k'} k' P(k')}=\frac{\langle k^2\rangle}{\langle k \rangle}.
\end{align}
The present paper is organized as follows. In section~\ref{sec: model} we will explain the model. From section~\ref{sec: dynamics of probability} to \ref{sec: diffusion}, we will derive the condition. In section~\ref{sec: intuition} we will give an intuition why $b/c>\langle k_\textrm{nn}\rangle$ is the condition. In section~\ref{sec: simulation} we will confirm the condition by numerical simulations. Finally, we will conclude in section~\ref{sec: conclusion}.
\section{Model}\label{sec: model}
\begin{figure}[htbp]
\begin{center}
\includegraphics[width=0.8\textwidth,keepaspectratio=true]{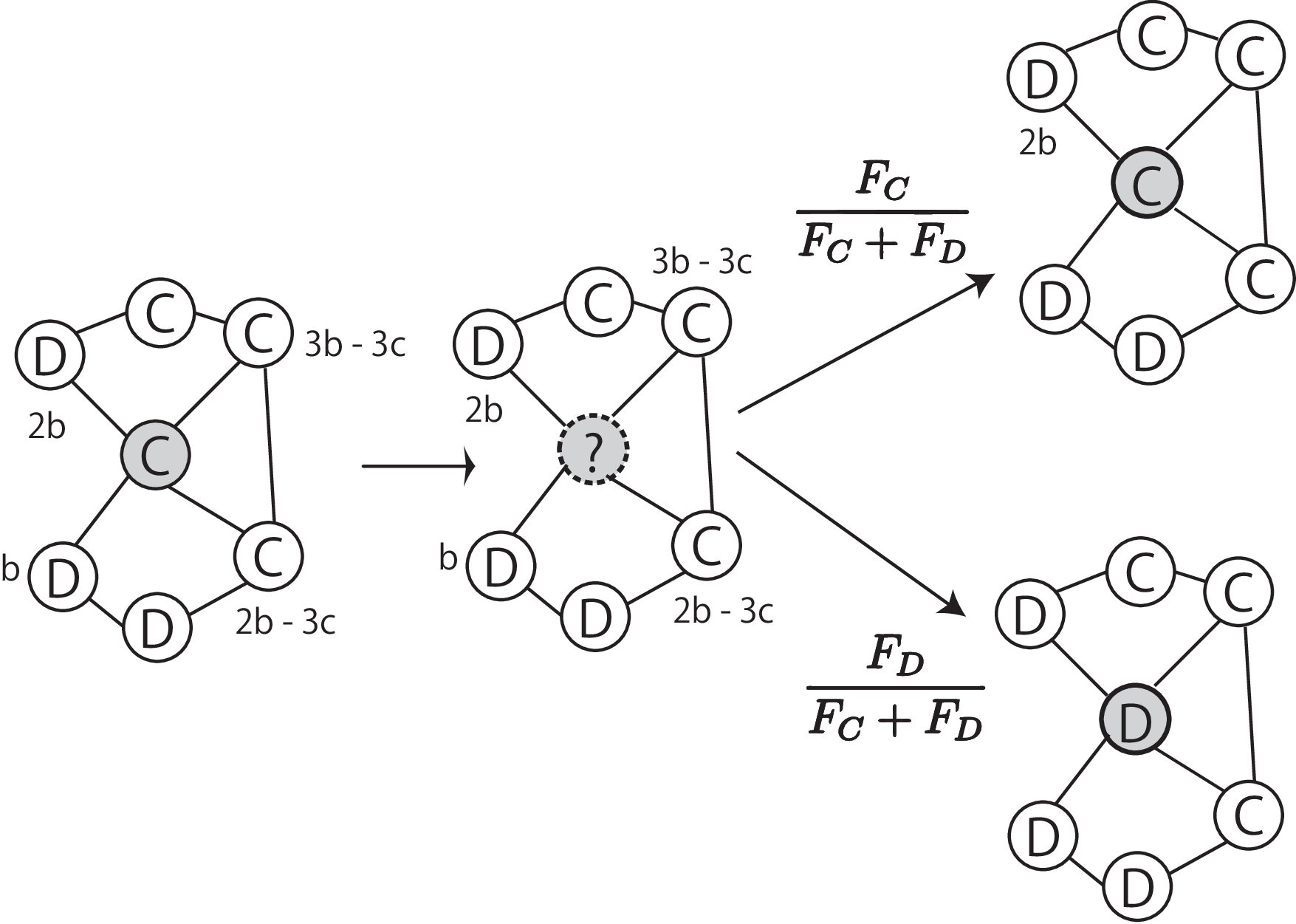}
\end{center}
\caption{\label{fig: rule of the game} An example of the update rule of the game on a network.
}\end{figure}

Let us introduce the model. We consider two kinds of players: cooperators (\textrm{C}) and defectors (\textrm{D}). Cooperators pay a benefit $b$ to their neighbors at a cost $c$, while defectors only receive the benefit. The game is a Prisoner's Dilemma (PD) game. The payoff matrix of the game is given by
\begin{align}
\bordermatrix{
      &\textrm{C} & \textrm{D}   \cr
    \textrm{C}& b-c & -c \cr
    \textrm{D}& b &  0 \cr
    }. \label{eq: payoff PD}
\end{align}
The game proceeds as follows. (See also Fig.~\ref{fig: rule of the game}.)

 At each time step, a randomly selected player dies. The adjacent players compete for the empty vertex, occupying it with a probability proportional to their fitness defined below; this is called the death-birth process. There is another interpretation of an evolutionary game \cite{smith1973logic}. At each time step, an player is randomly chosen and updates his strategy by imitating his neighbors' strategy with a probability proportional to their payoff. We can thereby regard the model as a model of human behavior, which is called imitation dynamics in social sciences.

The fitness of each player $F$ is given by the sum of constant term, the baseline fitness $1-w$ and the payoff from the game $P$ multiplied by $w$; namely, $F=1-w+wP$. We set $w\ll1$, which is called weak selection. Then the probabilities of strategies change only slowly.

In Fig.~\ref{fig: rule of the game}, for example, the total payoff of cooperators around the randomly chosen shaded vertex is $F_\textrm{C}=2-2w+w(5b-6c)$, while that of defectors is $F_\textrm{D}=2-2w+3wb$. The central randomly chosen player will be updated to a cooperator with the probability $F_\textrm{C}/(F_\textrm{C}+F_\textrm{D})$ and to be a defector with the probability $F_\textrm{D}/(F_\textrm{C}+F_\textrm{D})$.

 The update rule is a replicator dynamics extended to a network.  If a network is complete, it is equivalent to a replicator dynamics:
\begin{align}
\frac{d}{dt}p_i=wp_i(P_i-\bar{P}),
\end{align}
where $p_i$ is the probability of strategy $i$, $P_i$ is the mean payoff from the game of the player adopting strategy $i$, and $\bar{P}$ is the mean payoff from the game over all players.
This is called replicator equation on graphs \cite{ohtsuki2006replicator}.

In the following, we explain the criterion whether a network favors cooperation or not. As the initial condition,
we prepare a network in which all the vertices are occupied by defectors only.
Next, we replace one of them by a single cooperator and run the evolutionary games
until all the vertices are occupied either by only defectors, or by only cooperators.
There are only two terminal states: one is that the whole network is occupied by cooperators only and the other is that it is occupied by defectors only.
We iterate the same game a number of times and obtain the probability that only cooperators occupy the vertices.
This probability is called the fixation probability $\rho_\textrm{C}$~\cite{oxford2006evolutionary}.
If the selection neither favors nor opposes cooperation, the fixation probability $\rho_\textrm{C}$ is $1/N$, where $N$ is the network size, because the density of the cooperators in the initial condition is $1/N$.
If the fixation probability $\rho_\textrm{C}$ is larger than $1/N$, we say that the network structure favors cooperation, and vice versa.

We want to know the dynamics of the probability of cooperators $p_\textrm{C}$. However, we first study the more general problem of the following payoff matrix with strategy $A$ and $B$:
\begin{align}
\bordermatrix{
      &A &B   \cr
    A& x & y \cr
    B& z &  s \cr
    }.\label{eq:  general payoff}
\end{align}
A player is set on each vertex and plays games with its neighbors only.
We consider a network without degree-degree correlations. We want to know the dynamics of the probability distribution of cooperators $p_\textrm{C}$. However, we first study the problem of the general setting in Eq.~\eqref{eq:  general payoff}, and then of the Prisoner's Dilemma game.
Let $N(k)$ denote the number of vertices with degree $k$ and let $P(k)$ denote degree distribution. Let $P_\textrm{nn}(k_\textrm{nn})$ denote the probability that an adjacent vertex has degree $k_\textrm{nn}$; $P_\textrm{nn}(k_\textrm{nn})$ is the degree distribution of nearest neighbors, which is different from the degree distribution $P(k)$. Let $p_X(k)$ denote the probability of strategy $X$ on a vertex with degree $k$. We use the pair approximation~\cite{matsuda1992statistical}. Let $q_{X|Y}(k_\textrm{nn},k)$ denote the conditional probability of finding an $X$-player, given that the adjacent vertex is occupied by a $Y$-player, and that the degrees of the $X$-player and the $Y$-player are $k_\textrm{nn}$ and $k$, respectively. For example, the probability that a randomly chosen vertex is a $Y$-player with degree $k$ and a next neighboring vertex is a $X$-player with degree $k_\textrm{nn}$ is $p_Y(k)P(k)q_{X|Y}(k_\textrm{nn},k)P_\textrm{nn}(k_\textrm{nn})$ in pair approximation.

We now provide the outline of derivation. We need to know the dynamics of $p_A$, where $p_A$ is the probability of strategy $A$ in the whole network. For this purpose, we also need the dynamics of the conditional probabilities such as $q_{A|A}(k_\textrm{nn},k)$. As such, we first calculate $\dot{p}_A(k)$, and then calculate $\dot{q}_{A|A}(k_\textrm{nn},k)$. Then, we calculate the fixation probability $\rho_A$ by a diffusion approximation. The terminal state of the probability $p_A$ is either $1$ or $0$ only and the fixation probability of strategy $A$, $\rho_A$, is the probability that $p_A$ reaches $1$ in one game.
At the final stage, we substitute the payoff matrix~\eqref{eq:  general payoff} with the Prisoner's Dilemma game~\eqref{eq: payoff PD}, and then we have the condition of favoring cooperation $b/c>\langle k_\text{nn}\rangle$.

We will follow the derivation of Ref.~\cite{Ohtsuki.Nature.2006}. The difference is that we consider the heterogeneity of degree distribution explicitly. Therefore, a probability of strategy and a conditional probability depend on the degree $k$ and nearest neighbor degree $k_\textrm{nn}$, so that we write them as $p_X(k)$ and $q_{X|Y}(k_\textrm{nn},k)$ as defined above. In addition, in order to deal with a heterogenous network we use a mean-field approximation for network structure that we will explain in the following. As a result of the mean-field approximation we will have two kinds of the probability of strategy $p_X$ and $p_X^\textrm{nn}$, and two kinds of the conditional probability $q_{X|Y}$ and $q_{X|Y}^\textrm{nn}$. The definitions will be given below.

\begin{figure}[htbp]
\begin{center}
\includegraphics[width=.5\textwidth,keepaspectratio=true]{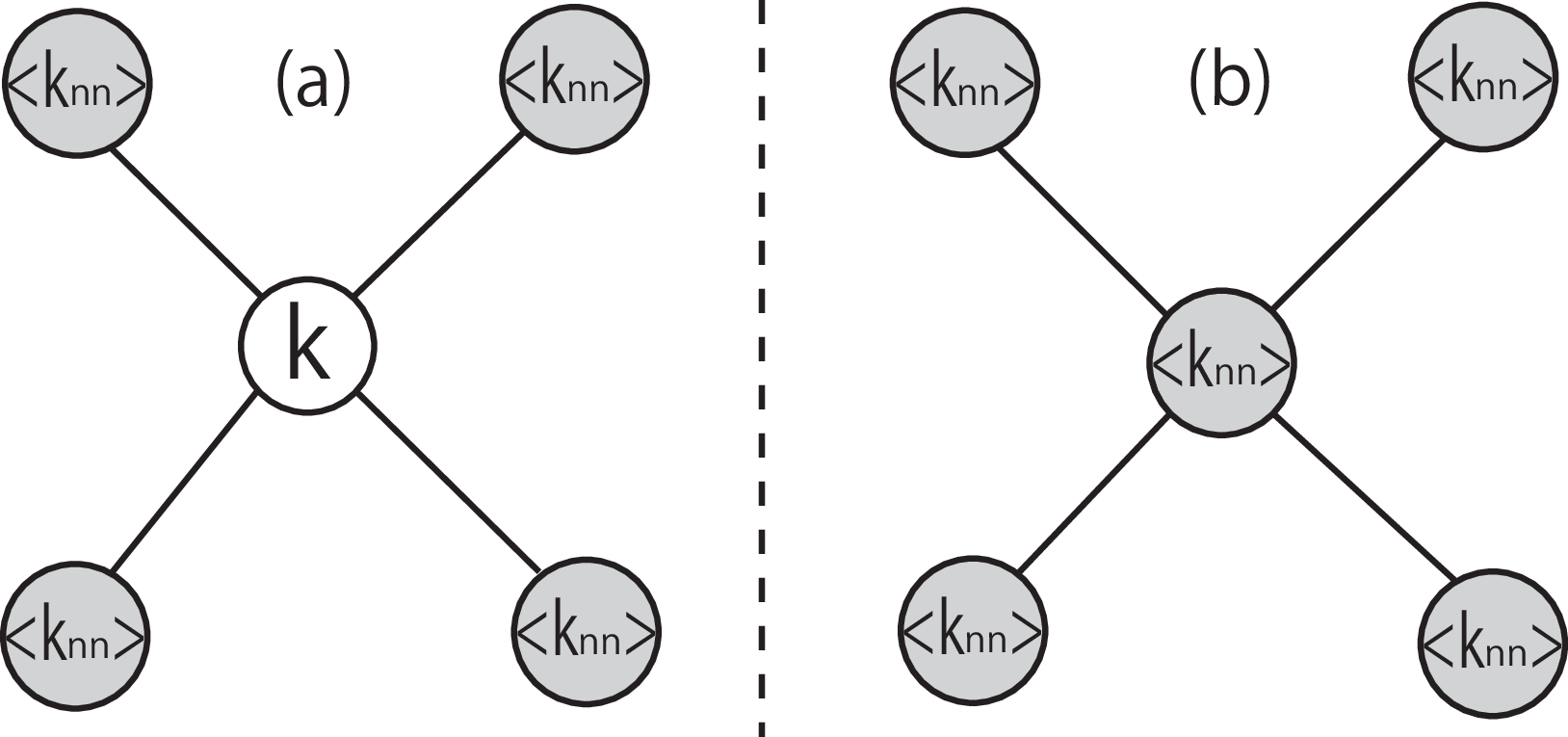}
\caption{\label{fig: intuition}: Mean-Field approximation. (a) A vertex with an arbitrary degree $k$ is surrounded by vertices with degree $\langle k_\textrm{nn}\rangle$. (b) A neighboring vertex with degree $\langle k_\textrm{nn}\rangle$ is also surrounded by vertices with degree $\langle k_\textrm{nn}\rangle$. }
\end{center}
\end{figure}

In the mean-field approximation for a heterogenous network structure, the degree of a vertex adjacent to any randomly chosen vertex is replaced by the mean degree of nearest neighbors $\langle k_\textrm{nn}\rangle$ as illustrated in Fig.~\ref{fig: intuition}(a). Subsequently, a vertex with degree $\langle k_\textrm{nn}\rangle$ is also surrounded by vertices with degree $\langle k_\textrm{nn}\rangle$ as illustrated in Fig.~\ref{fig: intuition}(b). Further, a degree of any randomly chosen vertex is replaced by the mean degree $\langle k\rangle$. Thus, in the mean-field approximation the probability of strategy $A$ on a randomly chosen vertex is given by $p_{A}(\langle k\rangle)$, and the probability of strategy $A$ on a neighbor of a randomly chosen vertex is given by $p_{A}(\langle k_\textrm{nn}\rangle)$. 

\section{Dynamics of probability of strategy}\label{sec: dynamics of probability}
First, we study the case where a $B$-player with degree $k$ is randomly chosen with probability $p_B(k) P(k)$ and changes the strategy from $B$ to $A$, and consequently $p_{A}$ increases by $1/N(k)$. Because we need $p_{A}(\langle k\rangle)$ and $p_{A}(\langle k_\textrm{nn}\rangle)$, we compute $p_{A}(k)$ first and then substitute $\langle k\rangle$ and $\langle k_\textrm{nn}\rangle$ later on.

Let $k_A$ and $k_B$ denote the number of $A$-players and $B$-players in the neighborhood of that randomly chosen $B$-player. As such, $k_A+k_B=k$.
Because of the mean-field approximation we replace the degree of any neighboring vertex by $\langle k_\textrm{nn}\rangle$, and the degrees of $A$-players and $B$-players around the randomly chosen $B$-player by both $\langle k_\textrm{nn}\rangle$. We define $q_{X|Y}^\textrm{nn}\equiv q_{X|Y}(\langle k_\textrm{nn}\rangle,\langle k_\textrm{nn}\rangle)$.
Degrees of neighboring vertices of the neighbors of any randomly chosen vertex are $\langle k_\textrm{nn}\rangle$ as illustrated in Fig.~\ref{fig: intuition}(b).
Thus, for example, the probability that a neighbor of an $A$-player neighboring to the randomly chosen vertex is a $B$-player is given by $q_{B|A}(\langle k_\textrm{nn}\rangle,\langle k_\textrm{nn}\rangle)\equiv q^\textrm{nn}_{B|A}$.
Therefore, the expected fitness of $A$-players and $B$-players adjacent to the chosen $B$-player, $f_{A}$ and $f_{B}$, respectively, are given by
\begin{align}
f_A&=1-w+w\left[(\langle k_\textrm{nn}\rangle-1)q_{A|A}^\textrm{nn}  x+\{(\langle k_\textrm{nn}\rangle-1)q_{B|A}^\textrm{nn}+1\}  y\right],\\
f_B&=1-w+w\left[(\langle k_\textrm{nn}\rangle-1)q_{A|B}^\textrm{nn}  z+\{(\langle k_\textrm{nn}\rangle-1)q_{B|B}^\textrm{nn}+1\}  s\right].
\end{align}
The probability that the randomly chosen $B$-player changes the strategy from $B$ to $A$ is then given by
\begin{align}
\frac{k_A f_A}{k_A f_A+k_B f_B}.
\end{align}
The probability that a randomly chosen $B$-player has $k_A$ $A$-players and $k_B$ $B$-players in the neighborhood is given by
\begin{align}
\sum_{k_A+k_B= k}\frac{k !}{k_A!k_B!}\left(q_{A|B}(\langle k_\textrm{nn}\rangle,k)\right)^{k_A}\left(q_{B|B}(\langle k_\textrm{nn}\rangle,k)\right)^{k_B}.
\end{align}
Therefore, the probability that a $B$-player is randomly chosen and changes the strategy from $B$ to $A$, and consequently $p_A(k)$ increases by $1/N(k)$ is given by
\begin{align}
\Pr\left(\triangle p_{A}(k)=\frac{1}{N(k)}\right)=&
p_B(k) P(k)\sum_{k_A+k_B= k}
\frac{\langle k\rangle!}{k_A!k_B!} \left(q_{A|B}(\langle k_\textrm{nn}\rangle,k)\right)^{k_A}\left(q_{B|B}(\langle k_\textrm{nn}\rangle,k)\right)^{k_B}
\notag\\
&\times\frac{k_A f_A}{k_A f_A+k_B f_B}. \label{eq: pa k plus}
\end{align}

Next, we consider the case where an $A$-player is randomly chosen and then the $A$-player changes the strategy from $A$ to $B$, and consequently $p_A(k)$ decreases by $1/N(k)$.

The probability that an $A$-player with degree $k$ is chosen is $p_A(k)P(k)$. Suppose that around the randomly chosen $A$-player, there are $k_{A}$ $A$-players and $k_{B}$ $B$-players and the probability that such a configuration occurs is given by
\begin{align}
\frac{ k! }{k_{A}!k_{B}! }\left(q_{A|A}(\langle k_\textrm{nn}\rangle,k)\right)^{k_{A}}  \left( q_{B|A}(\langle k_\textrm{nn}\rangle,k)\right)^{k_{B}}
\end{align}
The expected fitness of $A$-players and $B$-players adjacent to the randomly chosen $A$-player, $g_{A}$ and $g_{B}$, respectively, are given by
\begin{align}
g_A&=1-w+w\left[\{(\langle k_\textrm{nn}\rangle-1)q_{A|A}^\textrm{nn}+1\}  x+(\langle k_\textrm{nn}\rangle-1)q_{B|A}^\textrm{nn}  y \right],\\
g_B&=1-w+w\left[\{(\langle k_\textrm{nn}\rangle-1)q_{A|B}^\textrm{nn}+1\}  z+(\langle k_\textrm{nn}\rangle-1)q_{B|B}^\textrm{nn}  s\right].
\end{align}
Thus the probability that the randomly chosen $A$-player changes strategy from $A$ to $B$ is given by
\begin{align}
\frac{k_{B} g_{B}}{k_{A}g_{A}+k_{B}g_{B}}.
\end{align}

Therefore, the probability that an $A$-player is randomly chosen to update the strategy and then changes the strategy from $A$ to $B$, and consequently the probability of strategy $A$ with degree $k$ decreases by $1/N(k)$ is given by

\begin{align}
\Pr \left(\triangle p_{A}(k)=-\frac{1}{N(k)}\right)=&p_A(k)P(k)\sum_{k_{A}+k_{B}=k}\frac{k!}{k_{A}!k_{B}!}
\left( q_{A|A}(\langle k_\textrm{nn}\rangle,k)\right)^{k_{A}}  \left( q_{B|A}(\langle k_\textrm{nn}\rangle,k)\right)^{k_{B}}\notag\\
&\times\frac{k_{B} g_{B}}{k_{A}g_{A}+k_{B}g_{B}}. \label{eq: pa k minus}
\end{align}

Combining Eqs.~\eqref{eq: pa k plus} and \eqref{eq: pa k minus}, we have
\begin{align}
\frac{d}{dt}{p}_{A}(k)&=
\frac{1}{N(k)} \Pr\left(\triangle p_{A}(k)=\frac{1}{N(k)}\right)-\frac{1}{N(k)}\Pr\left(\triangle p_{A}(k)=-\frac{1}{N(k)}\right) . \label{eq: pak}
\end{align}
Because the degree of any randomly chosen vertex is $\langle k\rangle$, thus the probability of strategy $A$ in the network is $p_{A}(\langle k\rangle)$ in the mean-field approximation. Thus we have
\begin{align}
\frac{d}{dt}{p}_{A}=&\frac{1}{N(\langle k \rangle)}P(\langle k \rangle) p_B\sum_{k_A+k_B=\langle k \rangle} (q_{A|B})^{k_A}(q_{B|B})^{k_B}\frac{\langle k\rangle!}{k_A!k_B!}\frac{k_A f_A}{k_A f_A+k_Bf_B}\notag\\&-\frac{1}{N(\langle k \rangle)}P(\langle k \rangle)p_A
\sum_{k_A+k_B=\langle k\rangle}(q_{A|A})^{k_A}(q_{B|A})^{k_B}\frac{\langle k\rangle!}{k_A!k_B!}\frac{k_B g_B}{k_A g_A+k_B g_B}. \label{eq: mean prob}
\end{align}
Recall that $q_{X|Y}\equiv q_{X|Y}(\langle k_\textrm{nn}\rangle,\langle k\rangle)$.

We also need to have $E\left[\triangle p_{A}^\textrm{nn}\right]$, where $p_{X}^\textrm{nn}\equiv p_{X}(\langle k_\textrm{nn}\rangle)$ is the probability of strategy $A$ in the neighborhood of a randomly chosen player.
Substituting $\langle k_\textrm{nn}\rangle$ with $k$ of Eq.~\eqref{eq: pak}, we have
\begin{align}
\frac{d}{dt}{p}_A^\textrm{nn}=&\frac{1}{N(\langle k_\textrm{nn} \rangle)}P(\langle k_\textrm{nn} \rangle) p_B^\textrm{nn}\sum_{k_A+k_B=\langle k_\textrm{nn}\rangle} ({q^\textrm{nn}_{A|B}})^{k_A}  ({q^\textrm{nn}_{B|B}})^{k_B}\frac{\langle k_\textrm{nn}\rangle!}{k_A!k_B!}\frac{k_A f_A}{k_A f_A+k_Bf_B}\notag\\-&\frac{1}{N(\langle k_\textrm{nn} \rangle)}P(\langle k_\textrm{nn} \rangle)
p_A^\textrm{nn}\sum_{k_A+k_B=\langle k_\textrm{nn}\rangle} ({q_{A|A}^\textrm{nn}})^{k_A}  ({q_{B|A}^\textrm{nn}})^{k_B}\frac{\langle k_\textrm{nn}\rangle!}{k_A!k_B!}\frac{k_B g_B}{k_A g_A+k_B g_B}. \label{eq: mean pnn}
\end{align}
The fact that the set of equations is closed here is due to the mean-field approximation.

\section{Dynamics of conditional probability}

We then go on to the dynamics of conditional probabilities $q_{A|A}$ and $q_{A|A}^\textrm{nn}$.
We again use the mean-field approximation, replacing degree of any randomly chosen vertex by $\langle k\rangle$ and that of any adjacent vertex by $\langle k_\textrm{nn}\rangle$. First, we study the dynamics of $q_{A|A}\equiv q_{A|A}(\langle k_\textrm{nn}\rangle,\langle k\rangle)$.

We now study the case where $q_{A|A}$ increases.
For this purpose, suppose that a $B$-player is randomly chosen and then changes the strategy from $B$ to $A$.
Suppose that the randomly chosen $B$-player is linked to $k_{A}$ $A$-players and $k_{B}$ $B$-players. The probability that the $B$-player with this configuration is chosen is given by
\begin{align}
p_B P(\langle k\rangle)\frac{\langle k\rangle}{k_A!k_B!}(q_{A|B})^{k_A}(q_{B|B})^{k_B}.
\end{align}
The probability that the randomly chosen $B$-player changes the strategy from $B$ to $A$ is given by
\begin{align}
\frac{k_A f_A}{k_Af_A+k_Bf_B}.
\end{align}
After the $B$-player changes the strategy, the conditional probability $q_{A|A}$ increases by
\begin{align}
&\frac{p_A(\langle k\rangle,t)q_{A|A}(\langle k_\textrm{nn}\rangle,\langle k\rangle)N(\langle k\rangle)\langle k\rangle P_\textrm{nn}(\langle k_\textrm{nn}\rangle)+k_{A}}
{p_A(\langle k\rangle,t+\triangle t)N(\langle k\rangle)\langle k\rangle P_\textrm{nn}(\langle k_\textrm{nn}\rangle)}-q_{A|A}(\langle k_\textrm{nn}\rangle,\langle k\rangle),
\end{align}
because conditional probability $q_{A|A}(k_\textrm{nn},k)$ is given by the number of linked pairs of an $A$-player with degree $\langle k_\textrm{nn} \rangle$ and an $A$-player with degree $k$ divided by the number of linked pairs of an $A$-player with degree $k$ and a player of any strategy with degree $\langle k_\textrm{nn}\rangle$.
Using the fact that $p_A$ changes of order $O(w)$, which will be confirmed later, the above equation becomes
\begin{align}
\frac{k_A}{p_A N(\langle k\rangle)\langle k\rangle}+O(w).
\end{align}

Next, we are going to study the case where $q_{A|A}$ decreases.
Suppose that an $A$-player is randomly chosen and the $A$-player has $k_A$ $A$-players and $k_B$ $B$-players in the neighborhood. The probability that the $A$-player with such a configuration is chosen is
\begin{align}
p_A P(\langle k\rangle)\frac{\langle k\rangle}{k_A!k_B!}(q_{A|A})^{k_A}(q_{B|A})^{k_B}.
\end{align}
The probability that the $A$-player changes the strategy from $A$ to $B$ is
\begin{align}
\frac{k_B g_B}{k_A g_A+k_B g_B}.
\end{align}
After the $A$-player changes the strategy, the conditional probability $q_{A|A}$ decreases by
\begin{align}
\frac{k_A}{p_A N(\langle k\rangle)\langle k\rangle}+O(w). \label{eq: increase q}
\end{align}

Therefore, we have
\begin{align}
\frac{d}{dt}{q}_{A|A}=&\sum_{k_A+k_B=\langle k\rangle}\frac{k_A}{p_A \langle k \rangle N(\langle k\rangle)}p_B P(\langle k \rangle) \frac{\langle k\rangle!}{k_A!k_B!}(q_{A|B})^{k_A}(q_{B|B})^{k_B}\frac{k_A f_A}{k_A f_A +k_B f_B}\notag\\
&-\sum_{k_A+k_B=\langle k\rangle}\frac{k_A}{p_A \langle k \rangle N(\langle k\rangle)}p_A P(\langle k \rangle) \frac{\langle k\rangle!}{k_A!k_B!}(q_{A|A})^{k_A}(q_{B|A})^{k_B}\frac{k_B g_B}{k_A g_A +k_B g_B}\notag\\& +O(w) \label{eq: mean cond},
\end{align}
Analogously, we will compute $E[q_{A|A}^\textrm{nn}]$, where $q_{A|A}^\textrm{nn}\equiv q_{A|A}(\langle k_\textrm{nn}\rangle,\langle k_\textrm{nn}\rangle)$.

\begin{align}
\frac{d}{dt}{q}_{A|A}^\textrm{nn}=&\sum_{k_A+k_B=\langle k_\textrm{nn}\rangle}\frac{2k_A}{p_A^\textrm{nn} \langle k_\textrm{nn} \rangle N(\langle k_\textrm{nn}\rangle)}p_B^\textrm{nn} P(\langle k_\textrm{nn} \rangle)\frac{\langle k_\textrm{nn}\rangle!}{k_A!k_B!}({q_{A|B}^\textrm{nn}})^{k_A}({q_{B|B}^\textrm{nn}})^{k_B}\frac{k_A f_A}{k_A f_A +k_B f_B}\notag\\
-&\sum_{k_A+k_B=\langle k_\textrm{nn}\rangle}\frac{2k_A}{p_A^\textrm{nn} \langle k_\textrm{nn} \rangle N(\langle k_\textrm{nn}\rangle)}p_A^\textrm{nn} P(\langle k_\textrm{nn} \rangle)\frac{\langle k_\textrm{nn}\rangle!}{k_A!k_B!}({q_{A|A}^\textrm{nn}})^{k_A}({q_{B|A}^\textrm{nn}})^{k_B}\frac{k_B g_B}{k_A g_A +k_B g_B}\notag\\&+O(w) . \label{eq: qnn}
\end{align}
The derivation is given in detail in \ref{app: cond}.

\section{The System of Equations}
We now simplify the master equations in Eqs.~\eqref{eq: mean prob}, \eqref{eq: mean pnn}, \eqref{eq: mean cond}, and \eqref{eq: qnn}.
Transforming $\dot{p}_A$ in Eq.~(\ref{eq: mean prob}), we have
\begin{align}
\frac{d}{dt}{p}_A=\frac{\langle k\rangle-1}{\langle k\rangle N}p_{AB} \left( I_x x + I_b y - I_z z-I_s s\right)w +O(w^2) , \label{eq:dot pro}
\end{align}
 where
\begin{align}
\begin{split}
I_x\equiv(\langle k_\textrm{nn}\rangle-1)  q_{A|A}^\textrm{nn}  (q_{A|A}+q_{B|B})+q_{A|A},\\
I_y\equiv(\langle k_\textrm{nn}\rangle-1)  q_{B|A}^\textrm{nn}  (q_{A|A}+q_{B|B})+q_{B|B},\\
I_z\equiv(\langle k_\textrm{nn}\rangle-1)  q_{A|B}^\textrm{nn}  (q_{A|A}+q_{B|B})+q_{A|A},\\
I_s\equiv(\langle k_\textrm{nn}\rangle-1)  q_{B|B}^\textrm{nn}  (q_{A|A}+q_{B|B})+q_{B|B},
\end{split}
\end{align}
while the conditional probability $\dot{q}_{A|A}$ in Eq.~(\ref{eq: mean cond}) is transformed to
\begin{align}
\frac{d}{dt}{q}_{A|A}=\frac{1}{\langle k\rangle N p_A}p_{AB}\left[ 1+(\langle k\rangle-1)\{q_{A|B}-q_{A|A}\} \right]+O(w) .\label{eq:dot pair}
\end{align}
We confirmed that $\dot{p}_A$ is of order $O(w)$, whereas $\dot{q}_{A|A} $ is of order $O(w^0)$.
To derive Eq.~\eqref{eq:dot pro}, we used the mean-field relation
\begin{align}
q_{X|Y}p_Y=p_{XY}=p_{YX}=q_{Y|X}p_X; \label{eq: xy}
\end{align}
the reason why it holds is discussed in \ref{app: pair}. Owing to this relation the $O(w^0)$ terms in Eq.~\eqref{eq:dot pro} vanish.

Equations~\eqref{eq: mean pnn} and~\eqref{eq: qnn} lead to
\begin{align}
\frac{d}{dt}{p}_A^\textrm{nn}=\frac{\langle k_\textrm{nn}\rangle-1}{\langle k_\textrm{nn}\rangle N}p_{AB} \left( I_x^\textrm{nn} x + I_y^\textrm{nn} y - I_z^\textrm{nn} z-I_s^\textrm{nn} s\right)w +O(w^2) ,\label{eq:dot pro nn}
\end{align} where
\begin{align}
\begin{split}
I_x^\textrm{nn}\equiv(\langle k_\textrm{nn}\rangle-1)  q_{A|A}^\textrm{nn}  (q_{A|A}^\textrm{nn}+q_{B|B}^\textrm{nn})+q_{A|A}^\textrm{nn},\\
I_y^\textrm{nn}\equiv(\langle k_\textrm{nn}\rangle-1)  q_{B|A}^\textrm{nn}  (q_{A|A}^\textrm{nn}+q_{B|B}^\textrm{nn})+q_{B|B}^\textrm{nn},\\
I_z^\textrm{nn}\equiv(\langle k_\textrm{nn}\rangle-1)  q_{A|B}^\textrm{nn}  (q_{A|A}^\textrm{nn}+q_{B|B}^\textrm{nn})+q_{A|A}^\textrm{nn},\\
I_s^\textrm{nn}\equiv(\langle k_\textrm{nn}\rangle-1)  q_{B|B}^\textrm{nn}  (q_{A|A}^\textrm{nn}+q_{B|B}^\textrm{nn})+q_{B|B}^\textrm{nn},
\end{split}
\end{align}
 and
\begin{align}
\frac{d}{dt}{q}_{A|A}^\textrm{nn}&=\frac{2}{\langle k_\textrm{nn}\rangle N p_A^\textrm{nn}}p_{AB} \left[ 1+(\langle k_\textrm{nn}\rangle-1)\{q_{A|B}^\textrm{nn}-q_{A|A}^\textrm{nn}\} \right]+O(w) .\label{eq:dot pair nn}\
\end{align}
We thus confirmed that $\dot{p}_A^\textrm{nn}$ is of order $O(w)$ and $\dot{q}_{A|A}^\textrm{nn}$ is of order $O(w^0)$.
To derive Eq.~\eqref{eq:dot pro nn}, we used the relation $q_{X|Y}^\textrm{nn}p_Y^\textrm{nn}=p_{XY}=q_{Y|X}^\textrm{nn}p_X^\textrm{nn}$. Owing to this relation the $O(w^0)$ terms in Eq.~\eqref{eq:dot pro nn} vanish.

We now have the following system of equations:
\begin{align}
\begin{split}
\dot{p}_A&=F_1(p_A,q_{A|A},q_{A|A}^\textrm{nn})w+O(w^2),\\
\dot{p}_A^\textrm{nn}&=F_2(p^\textrm{nn}_A,q_{A|A}^\textrm{nn})w+O(w^2),\\
\dot{q}_{A|A}&=F_3(p_A,q_{A|A},q_{A|A}^\textrm{nn})+O(w),\\
\dot{q}_{A|A}^\textrm{nn}&=F_4(p^\textrm{nn}_A,q_{A|A}^\textrm{nn})+O(w),
\end{split}\label{eq: systme}
\end{align}
where the functions $F_i$ are defined in Eqs.~\eqref{eq:dot pro}, \eqref{eq:dot pro nn}, \eqref{eq:dot pair}, and \eqref{eq:dot pair nn}.
The equations of $\dot{p}$ and $\dot{p}^\textrm{nn}$ are of order $O(w)$, whereas the equations of $\dot{q}_{A|A}$ and of $\dot{q}_{A|A}^\textrm{nn}$ are of order $O(w^0)$.
Because $w\ll1$ meaning weak selection, the conditional probabilities $q_{A|A}$ and $q_{A|A}^\textrm{nn}$ converge to stationary values much faster than $p_A$ and $p_A^\textrm{nn}$ do. Thus the system very quickly converges to the slow manifold given by $F_3=0$ and $F_4=0$. Therefore, we assume that the following equations always hold:
\begin{align}
\begin{split}
1+(\langle k\rangle-1)&\{q_{A|B}-q_{A|A}\}=0,\\
1+(\langle k_\textrm{nn}\rangle-1)&\{q_{A|B}^\textrm{nn}-q_{A|A}^\textrm{nn}\}=0.
\end{split}\label{eq:slow manifold}
\end{align}

Because the relations concerning the strategy pair, $q_{X|Y}p_Y=p_{XY}=q_{Y|X}p_X$ and $q_{X|Y}^\textrm{nn}p_Y^\textrm{nn}=p_{XY}=q_{Y|X}^\textrm{nn}p_X^\textrm{nn}$, hold, the probabilities of strategies $p_A, p_B$, $p_A^\textrm{nn}$, and $p_B^\textrm{nn}$ and the conditional probabilities $q_{A|A},q_{B|A}$, $q_{B|B},q_{A|B}$, $q_{A|A}^\textrm{nn},q_{B|A}^\textrm{nn}$, $q_{B|B}^\textrm{nn}$, and $q_{A|B}^\textrm{nn}$ can be expressed in terms of $p_A, p_A^\textrm{nn}$, $q_{A|A}$, and $q_{A|A}^\textrm{nn}$. Further, using Eqs.~\eqref{eq:slow manifold}, we can express the conditional probabilities in terms of $p_A$ and $p_A^\textrm{nn}$. In other words, only $p_A$ and $p_A^\textrm{nn}$ are sufficient to express the other probabilities and conditional probabilities.

\section{The condition of favoring cooperation}\label{sec: diffusion}
Remember that our concern is the dynamics of $p_A$. We now approximate the dynamics of $p_A$ as a diffusion process~\cite{kimura1970ipg, van1992stochastic}. Eliminating the  conditional probabilities from Eqs.~\eqref{eq:slow manifold} and using Eqs.~(\ref{eq: pa k plus}, \ref{eq: pa k minus}), we have the expectation value of $\triangle p_A$ and the variance of $\triangle p_A$ as follows:
\begin{align}
E[\triangle p_A]=&\frac{1}{N(\langle k\rangle)}  \Pr\left(\triangle p_A=\frac{1}{N(\langle k\rangle)}\right)-\frac{1}{N(\langle k\rangle)}  \Pr\left(\triangle p_A=-\frac{1}{N(\langle k\rangle)}\right)\notag\\
\simeq& \frac{\langle k\rangle-2}{\langle k\rangle(\langle k\rangle-1)N}p_A(1-p_A) \left(\alpha p_A +\beta p_A^\textrm{nn}+ \gamma\right) w\triangle t \notag \\  \equiv& m(p_A)\triangle t ,\label{eq: drift}\\
\text{Var}[\triangle p_A]=&\left(\frac{1}{N(k)}\right)^2  \Pr\left(\triangle p_A=\frac{1}{N(\langle k\rangle)}\right)+\left(-\frac{1}{N(\langle k\rangle)}\right)^2  \Pr\left(\triangle p_A=-\frac{1}{N(\langle k\rangle)}\right)\notag\\
\simeq&\frac{2 (\langle k\rangle-2)}{(\langle k\rangle-1)N  N(\langle k\rangle)}p_A(1-p_A)\triangle t \equiv v(p_A)\triangle t ,
\end{align}
where
\begin{align}
&\alpha\equiv(x-y-z+s)(\langle k\rangle-2),\\
&\beta\equiv(x-y-z+s)\langle k\rangle(\langle k_\textrm{nn}\rangle-2) ,\\
&\gamma\equiv (x-y-z+s)+(x-y)\langle k \rangle+\langle k\rangle \langle k_\textrm{nn}\rangle (y-s).
\end{align}
The dynamics of $p_A$ is approximated by the diffusion process with the drift $m(p_A)$ and the variance $v(p_A)$ for unit time step $\triangle t$.

The fixation probability of strategy $A$, $\rho_A(r)$, for the initial probability \\$p_A(t=0)=r$, satisfies the following differential equation:
\begin{align}
0=m(r)\frac{d\rho_A(r)}{dr}+\frac{v(r)}{2}\frac{d^2\rho_A(r)}{dr^2} .\label{eq:diff eq}
\end{align}


Now, we use the Prisoner's Dilemma pay-off matrix given by
\begin{align}
\bordermatrix{
      &\textrm{C}    &\textrm{D}  \cr
    \textrm{C}& b-c & -c \cr
    \textrm{D}& b &  0 \cr
    }.\label{eq: pd payoff}
\end{align}
The differential equation~\eqref{eq:diff eq} becomes
\begin{align}
0=(b-c\langle k_\textrm{nn}\rangle)w \frac{d\rho_\textrm{C}(r)}{dr}+\frac{1}{N(\langle k \rangle)}\frac{d^2\rho_\textrm{C}(r)}{dr^2}.\label{eq:diff eq 2}
\end{align}
Since $w\ll1$, $\rho_\textrm{C}(1)=1$, and $\rho_\textrm{C}(0)=0$, we have the solution of Eq.\eqref{eq:diff eq 2} in the form
\begin{align}
\rho_\textrm{C}(r)\approx r+w \frac{N(\langle k\rangle)}{2 }(b-c\langle k_\textrm{nn}\rangle) r(1-r) .
\end{align}
As we discussed in section~\ref{sec: model}, the criterion that a network favors cooperation is $\rho_\textrm{C}(1/N)>1/N$.
Therefore, we have
\begin{align}
\frac{b}{c}>\langle k_\textrm{nn} \rangle  .\label{eq:condition for cooperation uncorrelated}
\end{align}
This is the condition that a network favors cooperation.
\section{Intuition: Why is $b/c>\langle k_\textrm{nn}\rangle$ the condition?}\label{sec: intuition}
\begin{figure}[htbp]
\begin{center}
\includegraphics[width=.25\textwidth,keepaspectratio=true]{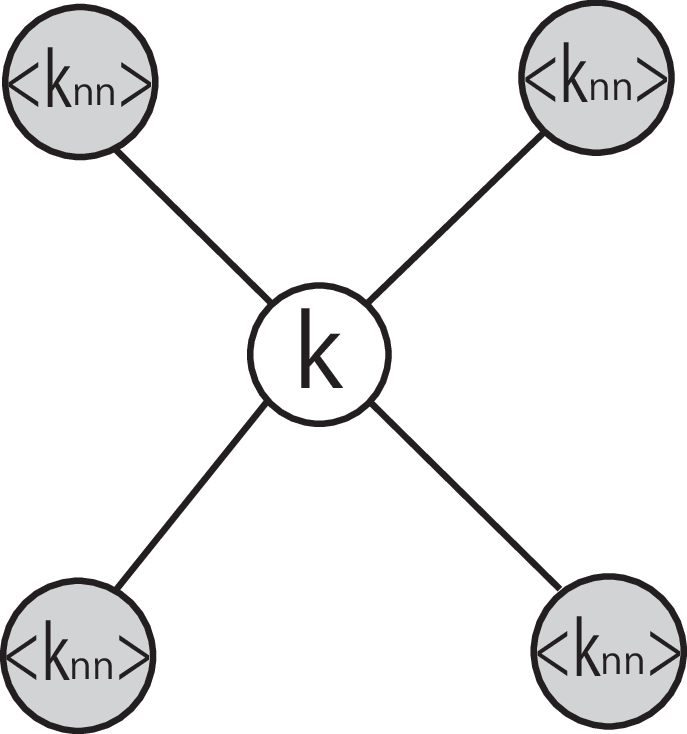}
\caption{\label{fig: intuition mod} In the mean-field approximation for a network structure, the degrees of vertices adjacent to any vertex are $\langle k_\textrm{nn}\rangle$.}
\end{center}
\end{figure}

We present intuitive reasoning of why the condition for favoring cooperation is $b/c>\langle k_\textrm{nn}\rangle$.
The point is that the mean degree of players competing for the vacant vertex is $\langle k_\textrm{nn}\rangle$.
In the mean-field picture, any vertex is surrounded by vertices with degree $\langle k_\textrm{nn}\rangle$, as illustrated in Fig.~\ref{fig: intuition mod}.

After we transform the second equation of Eq.~\eqref{eq:slow manifold}, replacing $A$ by \textrm{C} and B by \textrm{D}, and using $q_{\textrm{C}|\textrm{D}}^\textrm{nn}=1-q_{\textrm{D}|\textrm{D}}^\textrm{nn}$ obtained from $\dot{q}_{\textrm{C}|\textrm{C}}^\textrm{nn}=0$, the equation becomes
\begin{align}
(\langle k_\textrm{nn}\rangle-1)q_{\textrm{C}|\textrm{C}}^\textrm{nn}&=(\langle k_\textrm{nn}\rangle-1)p_\textrm{C}^\textrm{nn}+p_\textrm{D}^\textrm{nn},\\
(\langle k_\textrm{nn}\rangle-1)q_{\textrm{C}|\textrm{D}}^\textrm{nn}&=(\langle k_\textrm{nn}\rangle-1)p_\textrm{C}^\textrm{nn}-p_\textrm{C}^\textrm{nn}.
\end{align}
Subtracting the second equation from the first one, we have
\begin{align}
(\langle k_\textrm{nn}\rangle-1)(q_{\textrm{C}|\textrm{C}}^\textrm{nn}-q_{\textrm{C}|\textrm{D}}^\textrm{nn})=1 .\label{eq: intuition}
\end{align}

\begin{figure}[htbp]
\begin{center}
\includegraphics[width=.3\textwidth,keepaspectratio=true]{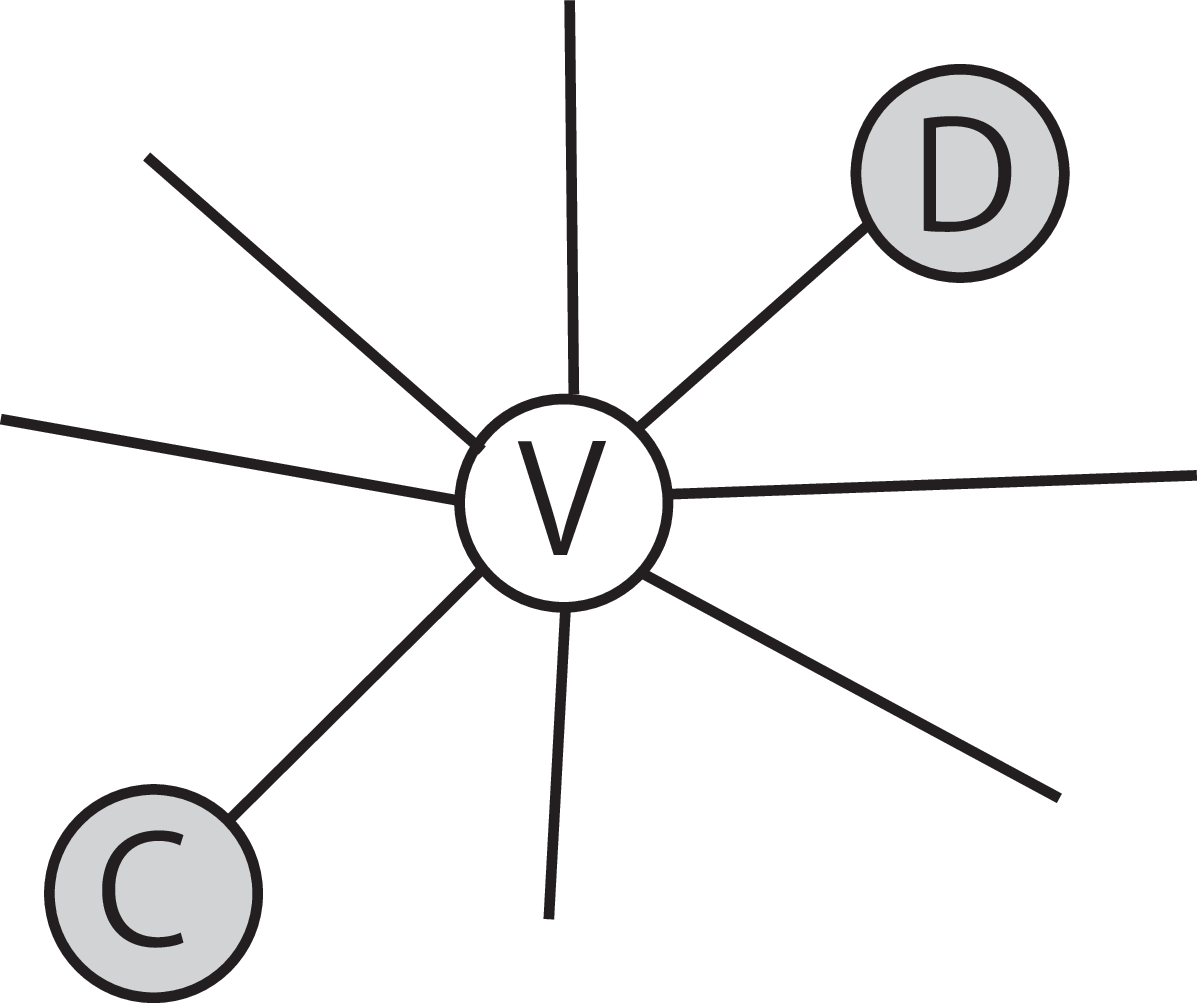}
\caption{\label{fig: compete}  A \textrm{C}-player and a \textrm{D}-player compete for a vacant vertex V. The degree of the vacant vertex is arbitrary. The mean degrees of the \textrm{C}-player and the \textrm{D}-player are both $\langle k_\textrm{nn}\rangle$.}
\end{center}
\end{figure}

Now, suppose that a \textrm{C}-player and a \textrm{D}-player compete for a vacant vertex V as illustrated in Fig.~\ref{fig: compete}. Because the payoff matrix of the Prisoner's Dilemma game is given by Eq.~\eqref{eq: pd payoff}, the expected payoff from the game for a \textrm{C}-player, $G_\textrm{C}^\textrm{nn}$, and that of a \textrm{D}-player, $G_\textrm{D}^\textrm{nn}$, are respectively, given by
\begin{align}
G_\textrm{C}^\textrm{nn}&=(b-c)q_{\textrm{C}|\textrm{C}}^\textrm{nn}(\langle k_\textrm{nn}\rangle-1)-c q_{\textrm{D}|\textrm{C}}^\textrm{nn}(\langle k_\textrm{nn}\rangle-1)+b\delta_{V,\textrm{C}}-c\\
G_\textrm{D}^\textrm{nn}&=b q_{\textrm{C}|\textrm{D}}^\textrm{nn}(\langle k_\textrm{nn}\rangle-1)+b\delta_{V,\textrm{C}},
\end{align}
where V is either C or D.
If $G_\textrm{C}^\textrm{nn}-G_\textrm{D}^\textrm{nn}>0$ holds, the network favors cooperation.
Using $(\langle k_\textrm{nn}\rangle-1)(q_{\textrm{C}|\textrm{C}}^\textrm{nn}-q_{\textrm{C}|\textrm{D}}^\textrm{nn})=1$, and $q_{\textrm{D}|\textrm{C}}^\textrm{nn}=1-q_{\textrm{C}|\textrm{C}}^\textrm{nn}$ in Eq.~\eqref{eq: intuition}, we have
\begin{align}
&G_\textrm{C}^\textrm{nn}-G_\textrm{D}^\textrm{nn}>0 ,\notag\\
\intertext{which yields}\quad &\frac{b}{c}>\langle k_\textrm{nn}\rangle .
\end{align}
Thus, if the condition $b/c>\langle k_\textrm{nn}\rangle$ is satisfied, cooperators are favored in the network.

\begin{figure}[htbp]
\begin{center}
\includegraphics[width=.45\textwidth,keepaspectratio=true]{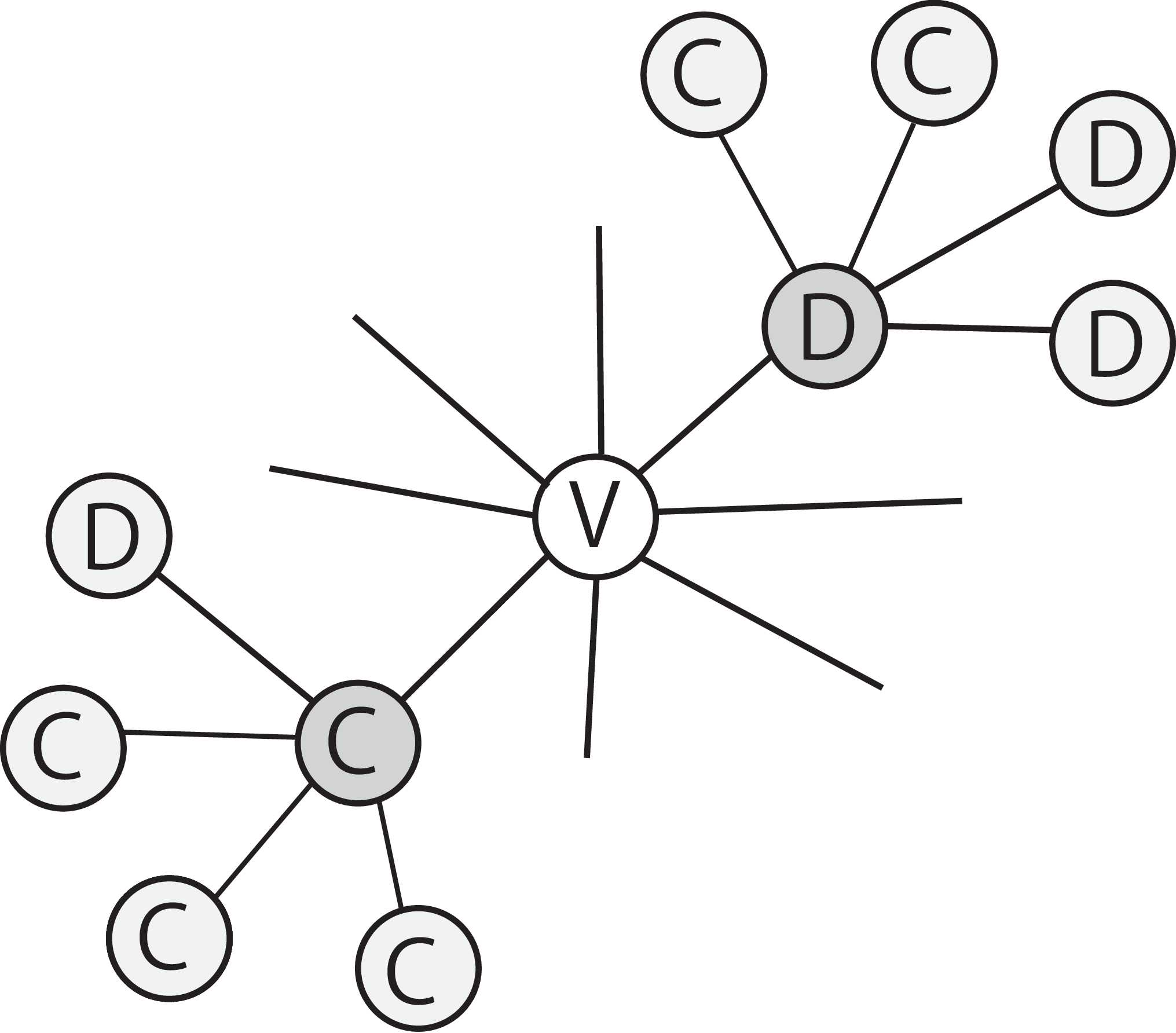}
\caption{\label{fig: compete intuition}  The \textrm{C}-neighbor has, on average, one more cooperator among their $\langle k_\textrm{nn}\rangle-1$ neighbors than the \textrm{D}-neighbor does.}
\end{center}
\end{figure}

In other words, we see from Eq.~\eqref{eq: intuition} that the \textrm{C}-neighbors of the vacant vertex V have, on average, one more \textrm{C} neighbor among their $\langle k_\textrm{nn}\rangle$ other neighbors than the \textrm{D}-neighbors of V do.  This extra benefit $b$ must outweigh the cost $c\langle k_\textrm{nn}\rangle$ incurred by the \textrm{C}-neighbors of V. Thus we must have $b/c > \langle k_\textrm{nn}\rangle$.
In the example illustrated in Fig.~\ref{fig: compete intuition}, the \textrm{C}-neighbor of the vacant vertex V has three cooperators as the neighbors, while the \textrm{D}-neighbor has two cooperators. The payoff from the game of the cooperator, $G_{\textrm{C}}=3b-c\langle k_\textrm{nn}\rangle$, must outweigh that of the defector, $G_{\textrm{D}}=2b$. Thus, we have the condition~\eqref{eq:condition for cooperation uncorrelated}.

\section{Which networks favor cooperation the most and the least}
Because our condition $b/c>\langle k_\textrm{nn}\rangle$ depends on $\langle k_\textrm{nn} \rangle$, we can see which of the three representative networks, regular, random, and scale-free, favors cooperation the most and the least. For this comparison purposes, we fix $\langle k \rangle(\equiv\mu)$ for the three networks. Even though $\langle k \rangle$ is the same, $\langle k_\textrm{nn} \rangle$ can be different.

In general, we have
\begin{align}
\langle k_\textrm{nn}\rangle=\frac{\langle k^2\rangle}{\langle k\rangle}=\frac{\sigma^2+\mu^2}{\mu} \label{eq: knn relation} .
\end{align}
where $\sigma^2$ is the variance of the degree distribution.
For a regular network, $\langle k_\textrm{nn}\rangle^\text{regular}=\mu$ since $\sigma^2=0$. The degree distribution of a random network is Poisson. Because the mean and the variance of Poisson distribution are the same, we have $\langle k_\textrm{nn}\rangle^\text{random}=\langle k\rangle+1$ for a random network.
Therefore, $\langle k_\textrm{nn} \rangle^\text{regular} < \langle k_\textrm{nn} \rangle^\text{random}$.
A scale-free network is a network in which the degree distribution $P(k)$ follows $P(k)\sim k^{-\gamma}$ typically with $2<\gamma\le3$. The mean degree of the nearest neighbors of a scale-free network of infinite size is
\begin{align}
\langle k_\textrm{nn}\rangle^\text{scale-free}=\frac{\langle k^2\rangle}{\langle k\rangle}\notag
=\left[ \int_1^\infty k^{2-\gamma}dk \right] \biggm/ \left[ \int_1^\infty k^{1-\gamma}dk \right]\Rightarrow \infty
\end{align}
for $2<\gamma\le3$.
Thus, the inequality $\langle k_\textrm{nn} \rangle^\text{regular} < \langle k_\textrm{nn} \rangle^{\text{random}} < \langle k_\textrm{nn} \rangle^{\text{scale-free}}$ holds for almost all the cases of interest.

Among the three network classes, a regular network favors cooperation the most and a scale-free network favors it the least. In an ideal scale-free network of infinite size and with $2<\gamma\le3$, cooperation is unfeasible because $\langle k_\textrm{nn} \rangle^\text{scale-free}$ is infinite. Because of Eq.~\eqref{eq: knn relation}, the network heterogeneity increases $\langle k_\textrm{nn}\rangle$. In other words, a heterogenous network suppresses cooperation.
The feature that the scale-free network suppresses cooperation is seen in the numerical simulations in Ref.~\cite{Ohtsuki.Nature.2006} and also agrees with Ref.~\cite{fu2009evolutionary}. However, Refs.~\cite{santos2005scale,santos2008social,santos2006evolutionary} claim that the scale-free network and the heterogenous network conversely favors cooperation. This is probably because the rule of the games are different. Whether or not the scale-free network and the heterogenous network favor cooperation depends on the details of the game, although it is occasionally believed that these favor cooperation irrespective of the rule of a game.
\section{If the number of players are small, then the cooperation is favored}
As is stated yet, scale-free networks are ubiquitously observed networks in the reality and cooperation is widely observed in nature. We show that cooperation is not favored in scale-free networks. You may think that they are contradicting. However, the key is the number of players. The mean degree of the nearest neighbors $\langle k_{nn}\rangle$ increases with the number of players in scale-free networks. Therefore, If the number of players are small, then the mean degree of nearest neighbors is small, and then the cooperation is favored. This would not be the case if the condition were determined by the mean degree $\langle k\rangle$, since the mean degree $\langle k\rangle$ is a fixed value determined by the exponent $\gamma$ when the network size $N$ is large enough. We are considering typcial scale-free networks by which we mean $2<\gamma\le3.$

\section{Simulations}\label{sec: simulation}

\begin{figure}[htbp]
\begin{center}
\includegraphics[width=\textwidth,keepaspectratio=true]{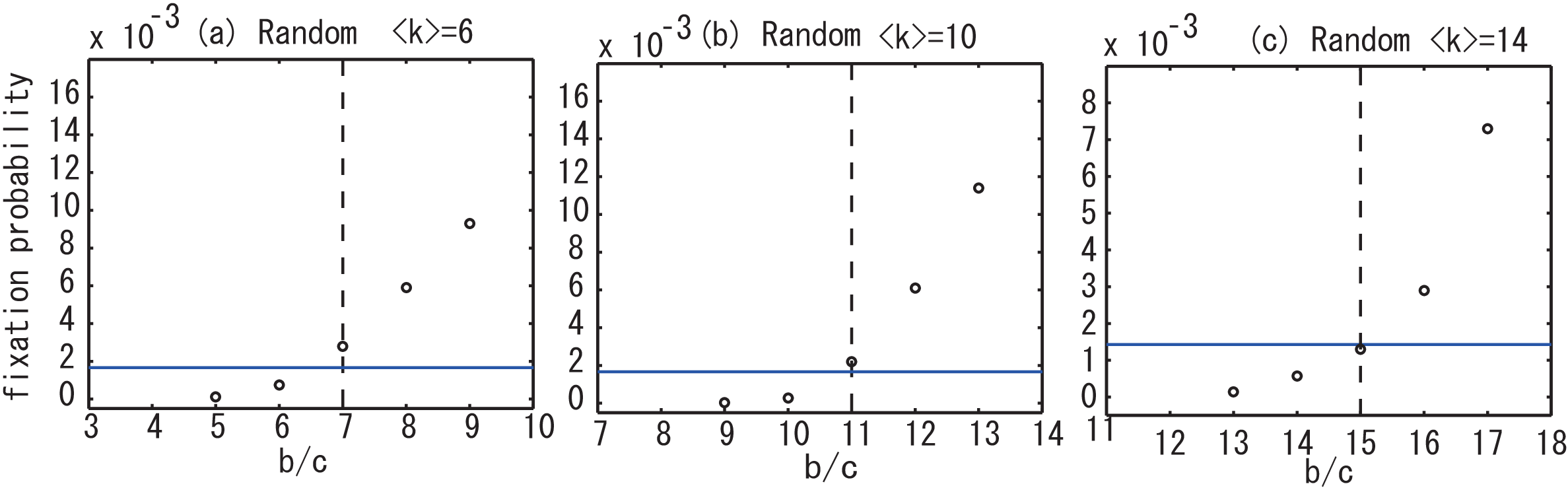}
\caption{\label{fig: numerical simulations 1} Results of numerical simulations for random networks. The horizontal straight line is the neutral probability $1/N$. The $x$ axis indicates $b/c$ while the $y$ axis indicates the fixation probability. The vertical broken line indicates the points where $b/c=\langle k_\textrm{nn}\rangle$. (a)--(c)  $\langle k \rangle=6, 10, 14$, the network size $N=600, 600, 700$, and $w=5\times10^{-3}, 7\times10^{-3}, 4\times10^{-3}$, respectively.}
\end{center}
\end{figure}

\begin{figure}[htbp]
\begin{center}
\includegraphics[width=0.7\textwidth,keepaspectratio=true]{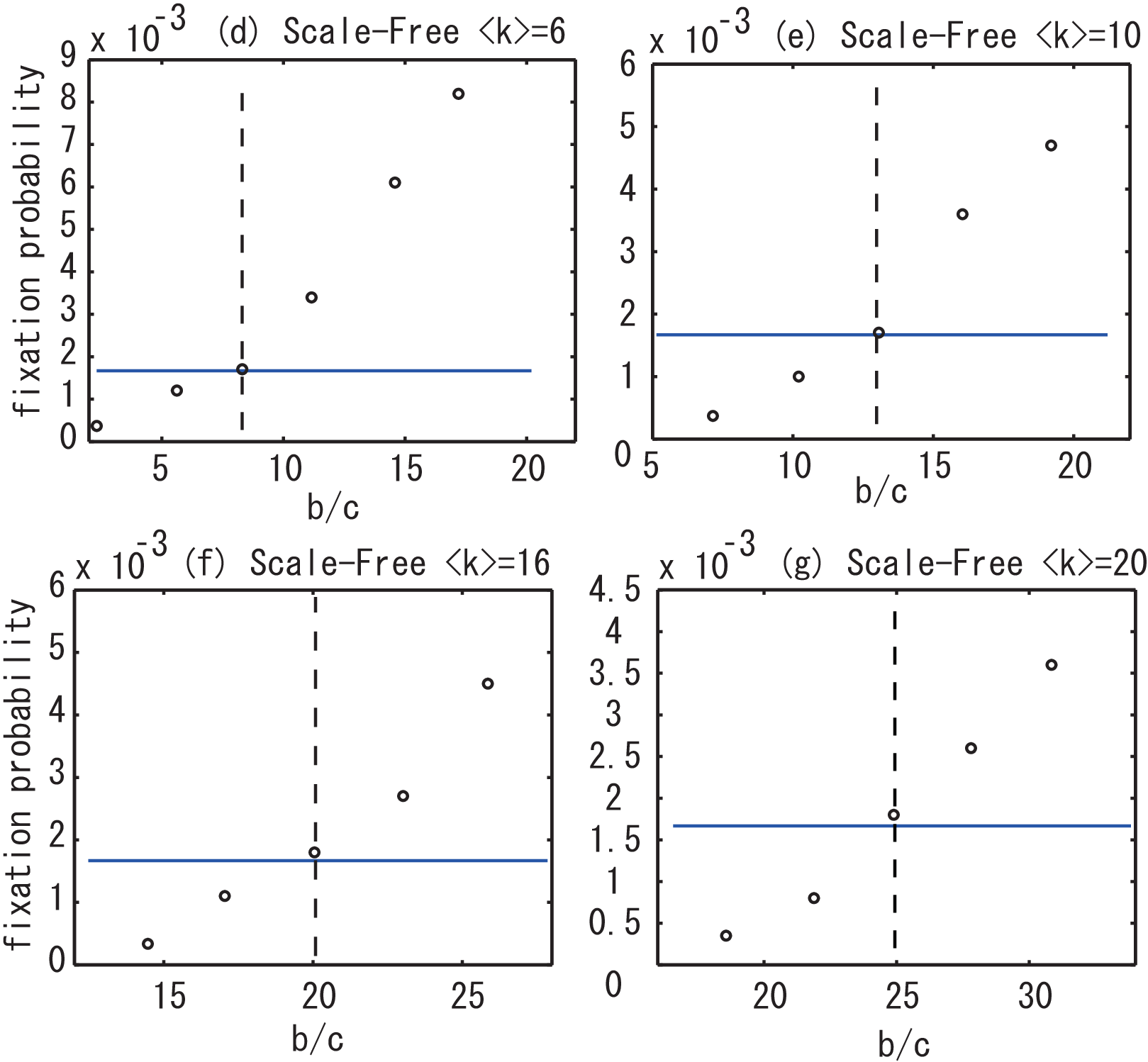}
\caption{\label{fig: numerical simulations 2} Results for numerical simulations of scale-free networks.
 (d)--(f)  $N=600$, $w=10^{-3}$, $\gamma=7.5$, and $\langle k \rangle=6, 10, 16, 20$, respectively.}
 \end{center}
\end{figure}



In the following, we show numerically that the condition (\ref{eq:condition for cooperation uncorrelated}) holds well, verifying our approximations. We simulate the game on several networks with $b/c$ set to different values.  The random networks are Erdos-Reyni random networks \cite{Erdos_Renyi_1959}. The scale-free networks are made by a preferential attachment mechanism \cite{Albert-L&aacute;szl&oacute;Barab&aacute;si10151999,dorogovtsev.pre.2000.85.21}. In Fig.~\ref{fig: numerical simulations 1} and Fig.~\ref{fig: numerical simulations 2}, the $x$ axis indicates $b/c$, while the $y$ axis does the fixation probability. The horizontal straight line corresponds to the neutral probability given by $1/N$. If a point is above the horizontal line, the network favors cooperation by definition; if the point is below the horizontal line, the network suppresses cooperation. The vertical broken line indicates the point where $b/c=\langle k_\textrm{nn}\rangle$. Our condition holds exactly if a point falls onto the crossing of the horizontal and vertical lines. Figure~\ref{fig: numerical simulations 1} and Fig.~\ref{fig: numerical simulations 2} show that for both random and scale-free networks, our condition holds well.

We constructed the scale-free networks in the following way.
First, we prepare a complete network consisting of $K$ vertices, in which all the vertices are connected to each other.  Next, a new vertex with $m$ links enters into the existing network. The probability that the new vertex is connected to an existing vertex $i$ is $\frac{k_i+A}{\sum_j( k_j +A)}$ where $k_i$ is the degree of vertex $i$. Next, another new vertex enters the exiting network in the same way. After repeating this process, we have a scale-free network. As $N$ gets large enough the exponent $\gamma$ of the degree distribution of the scale-free network asymptotically converges to $\gamma=3+A/m$.
In the simulation, we use $m=\langle k \rangle/2$, $K=m+1$, $N=600$, and $A$ is such that $3+A/m=7.5$, so that the exponent $\gamma$ is $7.5$.

\begin{figure}[htbp]
\begin{center}
\includegraphics[width=.55\textwidth,keepaspectratio=true]{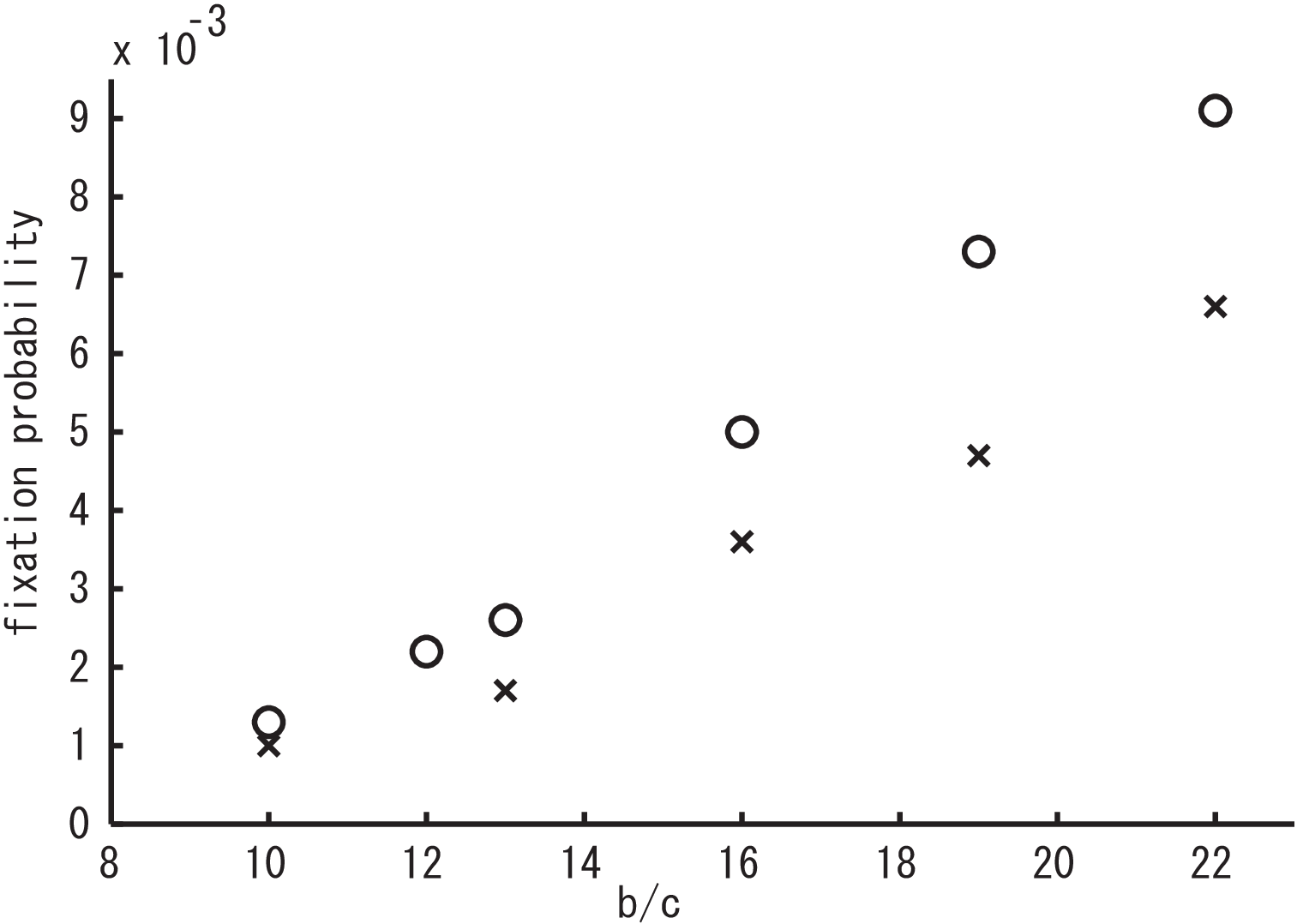}
\end{center}
\caption{Comparison between random networks and scale-free networks. The random network is indicated by circles; the scale-free network is indicated by crosses. The parameters are common: $\langle k \rangle=10$, $N=600$, and $w=10^{-3}$ for both networks. The exponent of scale-free network $\gamma$ is $7.5$. The $y$ axis indicates the fixation probability; the $x$ axis indicates $b/c$.}
\label{fig: compare}
\end{figure}

In Fig.~\ref{fig: compare}, we compare the random and the scale-free networks.
The network size $N$, $w$, and the mean degree $\langle k \rangle$ are taken to be the same.
The comparison shows that random networks favor cooperation more than
scale-free networks do, because the fixation probabilities for random networks are greater
for all the points.
\section{Conclusion}\label{sec: conclusion}
We study the Prisoner's Dilemma game where the payoff matrix is given by
\begin{align}
\bordermatrix{
      &\textrm{C}    &\textrm{D}  \cr
    \textrm{C}& b-c & -c \cr
    \textrm{D}& b &  0 \cr
    },
\end{align}
and analytically derive the condition of favoring cooperation for uncorrelated networks. The game is done under a death-birth process with weak selection. In summary, we obtained the following four results.

(i) Although it has been widely thought that $b/c>\langle k \rangle$ is the condition of favoring cooperation, we show that $b/c>\langle k_\textrm{nn} \rangle$ is the condition.
The mean degree of players competing for a vacant vertex is $\langle k_\textrm{nn}\rangle$ and the fitness of these adjacent players are determined by $\langle k_\textrm{nn}\rangle$; then, $\langle k_\textrm{nn}\rangle$ determines the outcome.

(ii) We show that among three representative networks, regular, random, and
scale-free, a regular network favors cooperation the most and a scale-free network the least. This is because the condition depends on the mean degree of nearest neighbors, $\langle k_\textrm{nn}\rangle$. Whereas the scale-free network has the largest mean degree of nearest neighbors, the regular network has the least for the same value of the mean degree $\langle k \rangle$.

(iii) In an ideal typical scale-free network characterized by the infinite number of vertices
with $\gamma\le3$, cooperation is unfeasible.

(iv) Although the scale-free network and the heterogeneous network favor cooperation in some cases, they suppress cooperation in our case. The scale-free network does not always favor cooperation irrespective of the game structure, although some occasionally believe so. Whether the scale-free network enhances or diminishes cooperation depends on details of the game.

(v) We show that if the number of players is small, then the cooperation is favored in scale-free networks. 

This would not be the case if the condition were determined by the mean degree $\langle k\rangle$, since the mean degree $\langle k\rangle$ is a fixed value determined by the exponent $\gamma$ when the network size $N$ is large enough.

\paragraph{Acknowledgement}\quad \\
The author would like to thank Z.R.~Struzik, N.~Hatano, H.~Ohtsuki, H.~Yoshikawa, M.~Kandori,  M.~Nakamaru, and two anonymous referees for their valuable comments, and Japan Society for the Promotion of Science for the financial support.

\appendix
\section{The derivation of $E[\triangle q^\textrm{nn}_{A|A}]$}\label{app: cond}
In the present appendix, we derive $E[\triangle q^\textrm{nn}_{A|A}]$ in Eq.~\eqref{eq: qnn}.
First, we are going to study the case where $q_{A|A}^\textrm{nn}$ increases.
Suppose that a $B$-player with degree $\langle k_\textrm{nn}\rangle$ that is linked to $k_A$ $A$-players and $k_B$ $B$-players in the neighborhood is randomly chosen. The probability that an $A$-player is chosen is given by

\begin{align}
p_B^\textrm{nn}P(\langle k_\textrm{nn}\rangle)
\frac{\langle k_\textrm{nn}\rangle}{k_A!k_B!}{q^\textrm{nn}_{A|B}}^{k_A}{q^\textrm{nn}_{B|B}}^{k_B},
\end{align}
where $p_X^\textrm{nn}\equiv p_X(\langle k_\textrm{nn}\rangle)$.
The probability that the chosen $B$-player changes the strategy from $B$ to $A$ is
\begin{align}
\frac{k_A f_A}{k_A f_A +k_B f_B}.
\end{align}
If the $B$-player changes the strategy from $B$ to $A$, the conditional probability $q_{A|A}^\textrm{nn}$ increases by
\begin{align}
\frac{2k_A}{p_A^\textrm{nn}N(\langle k_\textrm{nn}\rangle)\langle k_\textrm{nn}\rangle}+O(w),\label{eq: fraction}
\end{align}
where we have used the fact that $p_A^\textrm{nn}$ changes only of order $w$, which will be confirmed later.

\begin{figure}
  \begin{center}
  \includegraphics[width=\textwidth,keepaspectratio=true]{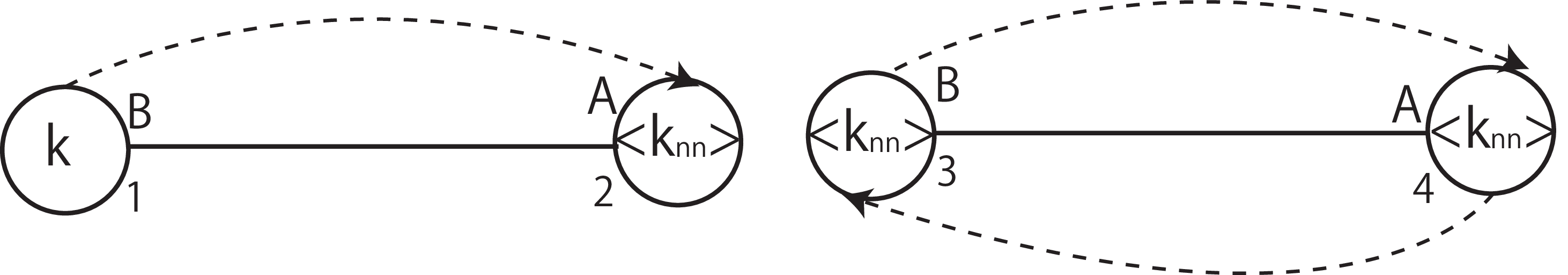}
  \caption{Pairs of an $A$-player and a $B$-player in two situations.}\label{fig: orientation}
  \end{center}
\end{figure}

Note the factor $2$ in the numerator of Eq.~\eqref{eq: fraction}. We are going to explain the reason of the factor in Fig.~\ref{fig: orientation}.
 The conditional probability increases when a $B$-player on the vertex 1 changes the strategy from $B$ to $A$ and the player is linked to an $A$-player on the vertex 2. The point is that when both of the degrees of the vertex 1 and the vertex 2 are $\langle k_\textrm{nn}\rangle$, the factor $2$ appears. Let A1 denote the strategy $A$ of the vertex 1 and A2 that of the vertex 2. Thus, both of the pairs A1--A2 and A2--A1 resulting from the change of strategy contribute to the increase in the conditional probability; one is seen from the vertex 1 and the other is seen from the vertex 2. When the vertex 1 was a $B$-player, on the other hand, the pair was B1--A1.
In the case of $q_{A|A}(\langle k_\textrm{nn}\rangle,\langle k\rangle)$, from the assumption that a randomly chosen vertex has the degree $\langle k \rangle$ and the vertices adjacent to the randomly chosen one has the degree $\langle k_\textrm{nn}\rangle$ the conditional probability would increase in one way seen from the vertex with the degree $\langle k\rangle$ in Eq.~\eqref{eq: increase q}.

Suppose that the vertex 1 changes the strategy from $B$ to $A$; then the pairs of $k$ and $\langle k_\textrm{nn}\rangle$ increases by one. On the other hand, suppose that the vertex 3 changes the strategy from $B$ to $A$; then the pairs of $\langle k_\textrm{nn}\rangle$ and $\langle k_\textrm{nn}\rangle$ increases by two. One is seen from the vertex 3 to 4 and the other is seen from the vertex 4 to 3.

Next, we are going to study the case where $q_{A|A}^\textrm{nn}$ decreases.
Suppose that an $A$-player with degree $\langle k_\textrm{nn}\rangle$ linked to $k_A$ $A$-players and $k_B$ $B$-players in the neighborhood is randomly chosen to update the strategy.
The probability that it happens is given by
\begin{align}
p_A^\textrm{nn}P(\langle k_\textrm{nn}\rangle)\frac{\langle k_\textrm{nn}\rangle}{k_A!k_B!}({q^\textrm{nn}_{A|A}})^{k_A}({q^\textrm{nn}_{B|A}})^{k_B}.
\end{align}
The probability that the randomly chosen $A$-player changes strategy from $A$ to $B$ is
\begin{align}
\frac{k_B g_B}{k_A g_A+k_B g_B}.
\end{align}
If the $A$-player changes the strategy from $A$ to B, then the conditional probability $q_{A|A}^\textrm{nn}$ increases by
\begin{align}
\frac{2k_A}{p_A^\textrm{nn}N(\langle k_\textrm{nn}\rangle)\langle k_\textrm{nn}\rangle}+O(w).
\end{align}

We thus have Eq.~\eqref{eq: qnn}:

\begin{align}
\frac{d}{dt} q_{A|A}^\textrm{nn}=&\sum_{k_A+k_B=\langle k_\textrm{nn}\rangle}\frac{2k_A}{p_A^\textrm{nn} \langle k_\textrm{nn} \rangle N(\langle k_\textrm{nn}\rangle)}p_B^\textrm{nn} P(\langle k_\textrm{nn} \rangle)\frac{\langle k_\textrm{nn}\rangle!}{k_A!k_B!}({q_{A|B}^\textrm{nn}})^{k_A}({q_{B|B}^\textrm{nn}})^{k_B}\frac{k_A f_A}{k_A f_A +k_B f_B}\notag\\
-&\sum_{k_A+k_B=\langle k_\textrm{nn}\rangle}\frac{2k_A}{p_A^\textrm{nn} \langle k_\textrm{nn} \rangle N(\langle k_\textrm{nn}\rangle)}p_A^\textrm{nn} P(\langle k_\textrm{nn} \rangle)\frac{\langle k_\textrm{nn}\rangle!}{k_A!k_B!}({q_{A|A}^\textrm{nn}})^{k_A}({q_{B|A}^\textrm{nn}})^{k_B}\frac{k_B g_B}{k_A g_A +k_B g_B}\notag\\&+O(w).
\end{align}

\section{The reason $q_{X|Y}p_Y=p_{XY}=p_{YX}=q_{Y|X}p_X$ holds}\label{app: pair}
In the present appendix, we argue the relation \eqref{eq: xy}.
The point is that in the mean-field approximation the degree of any randomly chosen vertex is $\langle k \rangle$, that of any neighboring degree is $\langle k_\textrm{nn}\rangle$, and that of a vertex attached to a randomly chosen link is also $\langle k_\textrm{nn}\rangle$. Note also that $q_{X|Y}$ is the conditional probability between randomly chosen vertex and its neighbor, and $q_{X|Y}^\textrm{nn}$ is the conditional probability between vertices on both ends of a randomly chosen link.

Pair probabilities are computed by two methods. In one method, we first choose a vertex randomly and check the strategy $X$ of the vertex. Then, we check the strategies of all vertices linked to the chosen vertex. The conditional probability of finding a strategy $Y$ on another vertex is $q_{Y|X}$. This procedure is carried out iteratively for all $N$ vertices. Because every link is connected to two vertices and every pair is counted twice, in this method, we count $2L$ pairs in total, where $L$ denotes the number of links in the network. The pair probabilities computed by this method let us understand the relation $q_{X|Y}p_Y=p_{XY}=q_{Y|X}p_X$.

In the other method of computing pair probability, we first choose a link and check the strategies of vertices of both ends of the chosen link. The probability of finding a strategy $X$ on one end is $p^\textrm{nn}_X$, and the conditional probability of finding a strategy $Y$ on the other end given that the strategy $X$ is already found is $q_{Y|X}^\textrm{nn}$. Then, we carry out this procedure iteratively for all $L$ links. In this method, we count $L$ pairs in total. The pair probabilities computed by this method let us understand the relation $q_{X|Y}^\textrm{nn}p_Y^\textrm{nn}=p_{XY}=q_{Y|X}^\textrm{nn}p_X^\textrm{nn}$. Because of the mean-field relation $q_{X|Y}p_Y=p_{XY}=q_{Y|X}p_X$, the $O(w^0)$ term in Eq.~\eqref{eq:dot pro} vanishes.

We also give another explanation of the relation $q_{X|Y}p_Y=q_{Y|X}p_X$.
The pair probability $p_{XY}$ is given by
\begin{align}
p_{XY}&=\frac{\text{\# of $XY$ pairs}}{\text{\# of Links}}\notag\\
&=\frac{\sum_{k_\textrm{nn},k}q_{X|Y}(k_\textrm{nn},k)P(k_\textrm{nn}|k)kNp(k)p_Y(k)}{N \langle k\rangle},
\end{align}
and the pair probability $p_{YX}$ is given by
\begin{align}
p_{YX}&=\frac{\text{\# of $YX$ pairs}}{\text{\# of Links}}\notag\\
&=\frac{\sum_{k_\textrm{nn},k}q_{Y|X}(k_\textrm{nn},k)P(k_\textrm{nn}|k)kNp(k)p_X(k)}{N \langle k \rangle}.
\end{align}
Further, the relation $p_{XY}=p_{YX}$ holds. 
Therefore, the following equation holds:
\begin{align}
&\frac{\sum_{k_\textrm{nn},k}q_{X|Y}(k_\textrm{nn},k)P(k_\textrm{nn}|k)kNp(k)p_Y(k)}{N \langle k\rangle}=p_{XY}\\
&=p_{YX}=\frac{\sum_{k_\textrm{nn},k}q_{Y|X}(k_\textrm{nn},k)P(k_\textrm{nn}|k)kNp(k)p_X(k)}{N \langle k \rangle} .\label{eq: 1}
\end{align}
Because of the mean-field approximation we replace the degree of nearest neighbors $k_\textrm{nn}$ by $\langle k_\textrm{nn}\rangle$ and the degree $k$ by $\langle k\rangle $:
\begin{align}
p_{XY}=&\frac{\sum_{k_\textrm{nn},k}q_{X|Y}(k_\textrm{nn},k)P(k_\textrm{nn},k)p(k)p_Y(k) k N }{N\langle k\rangle}\\
=&\frac{q_{X|Y}p_Y N\langle k\rangle }{N\langle k \rangle}=q_{X|Y}p_Y .
\end{align}
Therefore,  $p_{XY}=q_{X|Y}p_Y$ holds in the mean-field approximation.
Similarly, $p_{YX}=q_{Y|X}p_X$ holds.
Thus, $q_{Y|X}p_X=p_{YX}=p_{XY}=q_{X|Y}p_Y$ holds in the mean-field approximation.

\bibliographystyle{elsarticle-num}
\bibliography{bib_mean_field_cooperaton_on_non-regular_network}
\end{document}